\documentclass[11pt,showpacs]{revtex4-1}

\usepackage{epsfig}
\input epsf.tex
\usepackage{amssymb}
\usepackage{amsmath}
\usepackage{amsthm}
\usepackage{amsopn}

\def\comment#1{}

\def\beq{\begin{equation}}
\def\eeq{\end{equation}}
\def\bea{\begin{eqnarray}}
\def\eea{\end{eqnarray}}

\usepackage{ulem}
\usepackage{cancel}
\usepackage{color}

\begin{document}

\title{Using strong intense lasers to probe sterile neutrinos}

\author{Pouya Bakhti$^{1}$, Rohoollah Mohammadi$^{2}$\footnote{rmohammadi@ipm.ir} and She-Sheng Xue$^{3}$\footnote{xue@icra.it}}
\affiliation{$^{1}$ School of physics, Institute for research in fundamental sciences (IPM), Tehran, Iran.\\
$^{2}$ Iran Science and Technology Museum (IRSTM), Tehran, Iran.\\
$^{3}$
ICRANet, P.zza della Repubblica 10, I--65122 Pescara, and Physics Department, University of Rome {\it La Sapienza}, P.le Aldo Moro 5, I--00185 Rome, Italy.}

\begin{abstract}
A linearly polarized laser beam acquires its circular polarization by
interacting with a neutrino beam for the reason that the gauge-couplings of
left-handed neutrinos are parity-violated. Based on this phenomena, we
study the oscillations of active and sterile neutrinos in short baseline
neutrino experiments. Using the total fluxes of active and sterile neutrinos
in the $3+1$ framework, we show that the rate of generating circular polarization oscillates as a function of the distance $L$ neutrinos propagating from the source to the detector. By measuring such oscillation, one can possibly determine the mixing amplitudes of active and sterile neutrinos and their squared-mass difference. Moreover our proposal can constrain alternative scenarios such as Lorentz violation or quantum decoherence that explain short baseline neutrino anomalies.
\end{abstract}


\maketitle
\section{Introduction}
In the standard model (SM) of particle physics, three flavor eigenstates $\nu_
\alpha\,(\nu_e,\nu_\mu,\nu_\tau$) of active neutrinos participate in weak
interactions. The flavor eigenstates $\nu_\alpha\,(\nu_e,\nu_\mu,\nu_\tau$)
mix with the mass eigenstates $\nu_i\,(\nu_1,\nu_2,\nu_3)$
by a unitary matrix parametrized by three mixing angles
($\theta_{12},\theta_{13},\theta_{23}$) and one CP-violating phase $\delta$.
Several experiments studying solar, atmospheric and reactor active neutrinos
in past years provide strong evidences supporting the existence of active
neutrino oscillations \cite{osc1}, implying that active neutrinos are not
exactly massless although they chirally couple to gauge bosons.  (long-
baseline neutrino oscillations). The values of $\theta_{12}$, $\theta_{13}$,
$\theta_{23}$ and $\Delta m^2_{ij}$ are obtained from an up-to-date global
analysis \cite{grev} of solar, atmospheric, reactor and accelerator neutrino
data.

However, a few anomalous results
have emerged in short-baseline neutrino oscillation measurements and in cosmological data analysis,
which cannot be accommodated in the scheme of active neutrino oscillations with solar and atmospheric mass squared differences
(see for review \cite{rev}).
The most popular interpretation of such anomalies is based on a simple
extension of the scheme of active neutrino oscillations, involving new additional light sterile neutrinos with mass in the eV range \cite{sterile}.

The right-handed neutrino fields are fundamentally
different from the other elementary fermion fields because they are invariant
under the symmetries of the SM: they are singlets of $SU_c(3)\times SU_L(2)\times U_Y(1)$ gauge symmetries.
These right-handed neutrino fields are called sterile \cite{pont} because they
do not participate in weak, strong and electromagnetic interactions, and associate only with gravitational interaction.
By the theory, the number and mass of  right-handed
neutrino fields are not constrained. The essential characteristic of these
fields is that they are singlets under the SM symmetries, and hence sterile.
Therefore, the short-baseline experiments
(see for example \cite{Aguilar:2001ty,Aguilar-Arevalo:2013pmq,
Armbruster:2002mp,Giunti:2012tn}) are important for observing neutrino oscillations
with $E_\nu/L\sim\Delta m^2\sim {\rm eV^2}$.
This is crucial in the neutrino physics \cite{sterile} to check the existence
of sterile neutrinos, their mass scale and mixing with normal active neutrinos.

Recently it is shown that for the reason of active neutrinos being left-handed
and their gauge-couplings being parity-violated, linearly polarized laser
photons acquire their circular polarizations by interacting with neutrino beams
\cite{xue1}. This phenomenon can possibly be used to detect the fluxes  of active neutrinos, so as to gain some insight into the physics of active
and sterile neutrino oscillations.
In this Letter, we calculate the total flux of active and sterile neutrinos at
a fixed distance $L$ from the source to the detector of active neutrinos
(the source-detector distance in short), so as to determine the amplitude
$[\sin^2(2\theta_{\alpha s})]$ of the active neutrino $\nu_\alpha$
and the sterile neutrino $\nu_s$ oscillations, as well as the mass-squared difference
\begin{equation}\label{msds}
\Delta m^2=m^2_s-m^2_i
\end{equation}
where $m_s$ and $m_i$ indicate the masses of sterile and active neutrinos.

For alternative scenarios that explain short baseline neutrino anomalies, number of events of different baselines are the same while active sterile neutrino mixing predicts different number of events for different baselines. Therefore this proposal can also constrained alternative scenarios. Quantum decoherence\cite{Bakhti:2015dca}, CPT or Lorentz symmetries violation and mass varying neutrinos are some alternative scenarios that explain short baseline neutrino anomalies.

First we shortly recall the some theoretical and experimental results on the
short-baseline neutrino oscillations. After
discussing the time evaluation of the total fluxes of neutrino beams as a
function of the source-detector distance $L$, we calculate the generation of
circularly polarized laser beam due to the interaction of a linear
polarized laser beam with an active neutrino beam. We show how the mass-
squared difference $\Delta m^2$ of Eq.~(\ref{msds}) and the total fluxes of
active and sterile neutrinos can be possibly determined by measuring
the circular polarization of the laser beam interacting with active neutrino beam at the different source-detector distance $L$.

\section{theoretical and experimental aspects of
short-baseline neutrino anomaly}
According to the neutrino mixing hypotheses,
flavor neutrinos $\nu_l$ are superpositions of the mass eigenstates labeled with $\nu_i$ as follows [for example see \cite{book}]
\begin{equation}\label{mix0}
    \nu_l=\sum_{i=1}^{3+n_s}U_{li}\nu_i
\end{equation}
where $l=e,\mu,\tau$ and $i=1,2,3,...,n_s$. Here $n_s$ indicates the number of possible sterile neutrino species.
The active flavor neutrinos $\nu_l$ are identified by their charged current interactions with gauge bosons $W_{\mu}^\pm$.
\begin{equation}\label{lag0}
   \pounds_{\rm int} = \frac{g_w}{2\sqrt{2}}\sum_{l, i}
   \left[\bar{\psi}_{\nu_i}\gamma^{\mu}(1-\gamma^5)U_{i\,l}\psi_{l}W_{\mu}^+\,+
   \,\bar{\psi}_{l}\gamma^{\mu}(1-\gamma^5)U^\dagger_{li}\psi_{\nu_i}W_{\mu}^-\right],
\end{equation}
where the summations are over $l=e,\mu,\tau;\, i=1,2,3,\cdot\cdot\cdot n_s$.
In the case $n_s=1$, i.e., the $3+1$ model, the mixing matrix $U$ is represented as a $4\times4$ unitary matrix. In the following,
we adopt the parametrization used in Ref.~\cite{Bakhti:2013ora},
\begin{eqnarray}\label{unitary}
U\equiv\left( \begin{matrix}1 & 0\cr 0 & U_{\rm PMNS}\end{matrix}
\right)\cdot U_s
\end{eqnarray}
where $U_{\rm PMNS}$ is the standard $3\times 3$ PMNS matrix containing mixing elements between different flavors of active neutrinos, and $U_s$ represents a $4\times 4$ mixing matrix between the sterile neutrino and active neutrinos.
The mixing matrix $U_s$ is parameterized with rotation angles ($\gamma,\alpha, \beta$) and phase factors ($\delta_1,\delta_2$) \cite{Bakhti:2013ora}.
\comment{
\begin{eqnarray} \label{US}
U_s=\left( \begin{matrix} \cos \alpha  & \sin\alpha e^{i \delta_1} &
0 & 0\cr -\sin \alpha e^{-i \delta_1} &   \cos \alpha&0 & 0 \cr 0
 & 0 & 1 & 0 \cr 0 & 0 & 0 & 1\end{matrix} \right).
 \left( \begin{matrix} \cos \gamma  & 0 & \sin\gamma   & 0\cr 0 &   1 &0 & 0 \cr
-\sin\gamma
 & 0 & \cos\gamma  & 0 \cr 0 & 0 & 0 & 1\end{matrix} \right).
 \left( \begin{matrix} \cos \beta  &  0 & 0 & \sin\beta e^{i \delta_2} \cr 0 &  1 &0 & 0 \cr 0
 & 0 & 1 & 0 \cr - \sin\beta e^{-i \delta_2}& 0 & 0 & \cos \beta \end{matrix} \right)\ .
 \end{eqnarray}
}

Assuming that all CP-phases ($\delta_1,\delta_2$) are zero and $m_{s}\gg m_{1},m_{2},m_{3}$, therefore one has $\Delta m^{2}\equiv \Delta m^{2}_{si}\gg \Delta m^{2}_{12},\Delta m^{2}_{13}, \Delta m^{2}_{23} $ \cite{Giunti:2012tn}. Since the effects of oscillations depends on the sum of the mass-squared differences of two neutrino species, one keeps the leading term $\Delta m^{2}$ by neglecting  the small mass-squared differences $\Delta m^{2}_{12},\Delta m^{2}_{23}$ and $\Delta m^{2}_{13}$ for short baseline experiments \cite{book}.  As a result, the probabilities of an active $\alpha$-neutrino oscillating to another active $\beta$-neutrino, and to the sterile neutrino
$\nu_s$ can be written as follows (see for example Ref.~\cite{book})
\begin{eqnarray}
 P(\nu_{\alpha}\rightarrow\nu_{\beta})=\delta_{\alpha\beta}-\sin^{2}(\frac{\Delta m^{2}L}{4E_\nu})(U_{\alpha 0}U_{\beta 0}\sum_{i=1}^{3}U_{\alpha i}U_{\beta i}),\label{Pab}
\end{eqnarray}
and
\begin{eqnarray}
P(\nu_{\alpha}\rightarrow\nu_{s})=\sin^{2}(\frac{\Delta m^{2}L}{4E_\nu})(U_{\alpha 0}U_{0 0}\sum_{i=1}^{3}U_{\alpha i}U_{0 i}).\label{Pabs}
\end{eqnarray}
On the other hand, the oscillation
probability of the two-flavor $\nu_{\mu}\leftrightarrow\nu_{e}$ oscillation
is given by, (see for example Ref.~\cite{book})
\begin{eqnarray} \label{Pabe}
P(\nu_{\mu}\rightarrow\nu_{e})\approx \sin^{2}(2\theta_{\mu e})\sin^{2}(\frac{\Delta m^{2}L}{4E_\nu}).
\end{eqnarray}
Comparing Eq.~(\ref{Pab}) with Eq.~(\ref{Pabe}), one has
\begin{equation}\label{s-theta}
\sin^{2}(2\theta_{\mu e})=U_{\mu 0}U_{e 0}\sum_{i=1}^{3}U_{\mu i}U_{e i},
\end{equation}
where [$\sin^{2}(2\theta_{\mu e})$] is the amplitude of neutrino oscillation
[$\nu_{\mu}\rightarrow\nu_{e}$]. Analogously, one can obtain the amplitude of
electron disappearance [$\sin^{2}(2\theta_{e e})$] and the amplitude of muon
disappearance [$\sin^{2}(2\theta_{\mu\mu})$] in terms of the unitary $U$
matrix elements (\ref{unitary}). In addition, we  can
rewrite Eq.~(\ref{Pabs}) as
\begin{eqnarray}
P(\nu_{\alpha}\rightarrow\nu_{s})\approx\sin^{2}(2\theta_{\alpha s})\sin^{2}(\frac{\Delta m^{2}L}{4E_\nu})
,\label{Pabs1}
\end{eqnarray}
by defining the amplitude of neutrino oscillation
$\nu_{\alpha} \leftrightarrow \nu_s$
\begin{equation}\label{ofs}
\sin^{2}(2\theta_{\alpha s})=(U_{\alpha 0}U_{0 0}\sum_{i=1}^{3}U_{\alpha i}U_{0 i}).
\end{equation}

In the short baseline experiments for the two-flavor
$\nu_{\mu}\leftrightarrow\nu_{e}$ oscillation, the ranges of
$0.2\,{\rm eV^2}<\Delta m^{2}<10\,{\rm eV^2}$
and $0.01\,{\rm eV^2}<\Delta m^{2}<1\,{\rm eV^2}$
are respectively discussed by the LSND \cite{Aguilar:2001ty} and
MiniBooNE \cite{Aguilar-Arevalo:2013pmq} experiments.
Combining the results of the experiments KARMEN \cite{Armbruster:2002mp},
ICARUS \cite{Antonello:2012pq}, LSND and
MiniBooNE \cite{Aguilar-Arevalo:2013pmq} all together,
one obtains \cite{Palazzo:2013me,Giunti:2013aea}  $\Delta m^{2} \sim  0.5 {\rm eV^{2}}$ and
\begin{equation}\label{data}
    \sin^{2}2\theta_{\mu e} \sim 0.0015,\,\,\,\,\,\sin^{2}2\theta_{\mu\mu} \sim 0.03-0.05,\,\,\,\,\,\sin^{2}2\theta_{e e} \sim 0.093-0.13.
\end{equation}

Reactor experiments have played an important role in
the establishment of the short baseline oscillation. In short baseline
experiments at distances $L < 100$ m from the reactor core, at ILL-Grenoble,
Goesgen, Rovno, Krasnoyarsk, Savannah River and
Bugey \cite{reactor}, the measured rate of $\bar{\nu}_e$ was found to be
in a reasonable agreement with that predicted from the reactor anti neutrino
spectra, though slightly lower than theoretically expected, with the measured/
expected ratio at $0.976\pm 0.024$. Neutrino oscillations in
distances $L < 100$ m from the reactor core are not expected in the
theoretical scenario of flavor mixing of three active neutrinos without sterile neutrinos.
In the short baseline experiments based on these reactors,
the deviation of experimental data from theoretical expectations
is called the reactor anomaly \cite{Mention:2011rk}.
Another deviation of experimental data of neutrino fluxes from
theoretical estimations based on the flavor mixing of three
active neutrinos is related to the GALLEX and SAGE experiments.
This deviation is called the gallium anomaly
\cite{Giunti:2012tn,Abdurashitov:2005tb}. Both the reactor
and gallium anomalies can possibly be interpreted in the 3+1 framework with
$\Delta m^{2} \gtrsim  1 \,{\rm eV^{2}} $ and $\sin^{2}2\theta_{ee} \sim 0.17$
\cite{Palazzo:2013me,Abazajian:2012ys}.
In Ref.~\cite{Giunti:2013aea}, the global analysis of the short baseline
neutrino oscillation in the scheme of the 3+1 neutrino mixing determines
$0.82\,{\rm eV^2}<\Delta m^{2} < 2.19\, {\rm eV^2}$ at the 3$\,\sigma$-level, and the assumption of no oscillation between flavor and sterile neutrinos
is excluded at 6$\,\sigma$-level.

The consideration of two or more sterile neutrino mass eigenstate also are used to explain short baseline neutrino anomalies where the experiment could also constrain these scenarios too. Active sterile mixing predicts different number of events for different baseline. Other scenarios explain these anomalies without consideration sterile neutrino, predicts no difference between difference baselines. These alternative scenarios explain flavor changes by modifying quantum such as quantum decoherence, or consideration of additional effective Lagrangian or effect of environment causes flavor change in short baselines. These flavor changes occurred between active flavors thus flux of active flavors neutrinos does not change. Establishment of sterile neutrino versus other alternatives is important from phenomenological point of view and could be test via this proposal because the effect is approximately the same for all active flavors.

\section{The Time Evaluation of Active Neutrino Flux }
Suppose that (i) the initial total energy flux
$F_{\nu}(0)$ of active neutrinos produced from the source consists of
the electron neutrino flux $F_{\nu_e}=\mathcal{C} F_{\nu}(0)$ and
muon neutrino flux
$F_{\nu_\mu}=(1-\mathcal{C})F_{\nu}(0)$, where $\mathcal{C} <1$; (ii)
there is the oscillation between the active neutrinos and light sterile neutrinos, this means that total flux of active neutrinos is a
function of the source-detector distance $L$ in the short baseline experiments. Thus we can write the total flux of active neutrinos as
\begin{eqnarray}\label{flux1}
    F_{\nu}(L)&=&F_{\nu}(0)-c\int\frac{d^3q}{(2\pi)^3}\left[ q\,f_{\nu_e}P_{\nu_{e}\rightarrow\nu_{s}}+
    q\, f_{\nu_\mu}P_{\nu_{\mu}\rightarrow\nu_{s}}\right]\nonumber\\
		&=& F_{\nu}(0)-F_{\nu_s}(L),
\end{eqnarray}
where $E_\nu\simeq q_\nu$, $f_{\nu_e,\nu_\mu}$ are the
energy-distribution functions of neutrino beams,
and the second term on the right-handed side
indicates the total flux $F_{\nu_s}$ of sterile neutrinos, which propagates
approximately in the speed of light $c$ for
$E_{\nu_s}\simeq q_{\nu_s}\gg m_{\nu_s}$. For scenarios without consideration of sterile neutrino $P_{\nu_{a}\rightarrow\nu_{s}}=0$ where $a=e,\mu$, thus flux of active neutrinos does not change $(F_{\nu}(L)=F_{\nu}(0)$).

On the other hand,
because electron and muon neutrinos are oscillating each other while they are propagating,
the fluxes $F_{\nu_e}(L)$ and $F_{\nu_\mu}(L)$ of electron and muon
neutrinos are related each other as a function of
the source-detector distance $L$. Using Eq.~(\ref{Pab}), we write the
total energy flux of the active flavor neutrinos as
\begin{eqnarray}\label{flux0}
F_{\nu}(L)&=& F_{\nu_e}(L)+F_{\nu_\mu}(L)\nonumber\\
&=& c\int\frac{d^3q}{(2\pi)^3}\Big[ (P_{\nu_{e}\rightarrow\nu_{e}}+P_{\nu_{e}\rightarrow\nu_{\mu}}+P_{\nu_{e}\rightarrow\nu_{\tau}}) q\,f_{\nu_e}\nonumber\\
&+&
(P_{\nu_{\mu}\rightarrow\nu_{\mu}}
+P_{\nu_{\mu}\rightarrow\nu_{e}}+P_{\nu_{\mu}\rightarrow\nu_{\tau}})q\, f_{\nu_\mu}\Big].
\end{eqnarray}
Suppose that the initial neutrino beam has a very narrow
energy distribution around the average neutrino energy $\bar{E}_\nu$,
and a very small angular distribution (divergence angle) that can be negligible.
Using Eq.~(\ref{Pabs}), we write the neutrino energy fluxes (\ref{flux0}) as follows,
\begin{eqnarray}
 F_{\nu}(L) &=& F_{\nu_e}(0)\left( 1+[-\sin^{2}(2\theta_{e e})+\sin^{2}(2\theta_{e \mu})+\sin^{2}(2\theta_{e \tau})]\sin^{2}(\frac{\Delta m^{2}L}{4\bar{E}_\nu})\right) \nonumber\\
 &+& F_{\nu_\mu}(0)\left( 1+[-\sin^{2}(2\theta_{\mu \mu})+\sin^{2}(2\theta_{\mu e})+\sin^{2}(2\theta_{\mu \tau})]\sin^{2}(\frac{\Delta m^{2}L}{4\bar{E}_\nu})\right),\label{flux2}
\end{eqnarray}
which are only functions of neutrino average energy $\bar{E}_\nu$ and the source-detector distance $L$.

Using Eqs.~(\ref{flux1}), (\ref{flux0}) and (\ref{flux2}), we approximately
obtain the total
sterile neutrino flux
\begin{eqnarray}
F_{\nu_s}(L) &\approx & F_{\nu}(0)\Big[\mathcal{C}\left(\sin^{2}(2\theta_{e e})-\sin^{2}(2\theta_{ e\mu})-\sin^{2}(2\theta_{e\tau})\right)\nonumber\\
  &+& (1-\mathcal{C})\left(\sin^{2}(2\theta_{\mu \mu})-\sin^{2}(2\theta_{ \mu e})-\sin^{2}(2\theta_{\mu\tau})\right)\Big]\sin^{2}(\frac{\Delta m^{2}L}{4\bar{E}_\nu})\nonumber\\
  &=&F_{\nu}(0)\Big[\mathcal{C}\sin^{2}(2\theta_{e s})+(1-\mathcal{C})\sin^{2}(2\theta_{\mu s})\Big]\sin^{2}(\frac{\Delta m^{2}L}{4\bar{E}_\nu}), \label{flux3}
\end{eqnarray}
where we define
\begin{eqnarray}
 \sin^{2}(2\theta_{\mu s})&\equiv&\sin^{2}(2\theta_{\mu \mu})-\sin^{2}(2\theta_{ \mu e})-\sin^{2}(2\theta_{\mu\tau}),  \nonumber\\
   \sin^{2}(2\theta_{e s})&\equiv&\sin^{2}(2\theta_{e e})-\sin^{2}(2\theta_{ e \mu})-\sin^{2}(2\theta_{e \tau}), \label{sin}
\end{eqnarray}
i.e., the mixing angles between active and sterile neutrinos
are expressed in terms of the mixing angles between different active
neutrinos.
These results of Eqs.~(\ref{flux1}), (\ref{flux2}) and (\ref{flux3}) show that the the total fluxes of active and sterile neutrinos
oscillate each other while they are propagating from the
active neutrino source to detector in the short-baseline experiments.

\section{Oscillation of circular polarization of laser beam}
As discussed in Ref.~\cite{xue1}, for the reason of active neutrinos being left-handed and their gauge-couplings
being parity-violated, the circular polarization of laser photons
is generated by interactions between a linearly polarized laser beam with an active neutrino beam.
The intensity of circular polarization represented by the
Stock V-parameter is expressed in terms of the intensity $Q$
of linearly polarized laser beam and the total flux $F_{\nu}(L,\bar{E}_\nu)$ of the active Dirac neutrino beam \cite{xue1}
\begin{eqnarray}
 \frac{\Delta V}{Q} 
\approx  2.37\cdot 10^{-36} ({\rm cm}^2) \left(\frac{ F_{\nu}(L,\bar{E}_\nu)}{k}\right)\,\Delta t,
  \label{circular1}
\end{eqnarray}
where $k$ is the mean energy of laser photons,
$\Delta t$ is the interacting time
of laser beams with the active neutrino beam at the source-detector distance
$L$. Note that we do not consider here the
circular polarization produced due to the interaction of laser beam directly
with sterile neutrinos, for the reason that laser photons
coupling to sterile neutrinos is assumed to be very small,
compared with their coupling to active neutrinos. The presence of
active and sterile neutrinos mixing and the oscillation leads to the  oscillating nature of the total flux $F_{\nu}(L,\bar{E}_\nu)$
of active neutrinos. As a result, Eq.~(\ref{circular1})
clearly shows that the circular polarization
$ \Delta V/Q$ oscillates as a function of the source-detector distance
$L$.

\begin{figure}
  \includegraphics[width=4in]{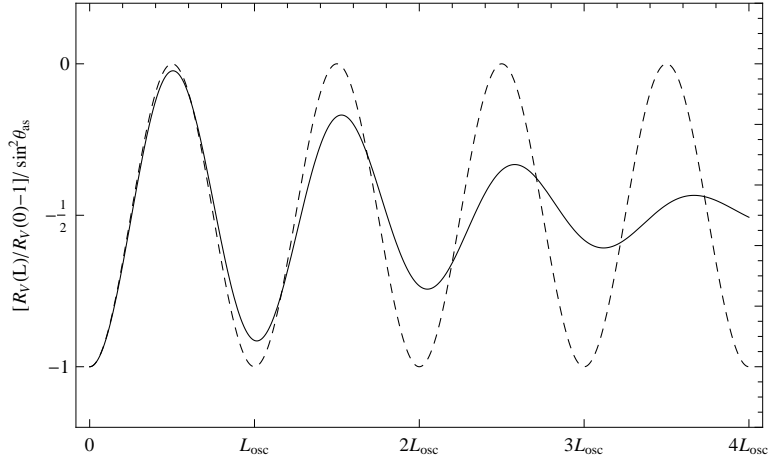}\\
  \caption{This plot shows the oscillation of $(R_{_{\rm V}}(L)-R_{_{\rm V}}(0))/R_{_{\rm V}}(0)\sin^{2}2\theta_{ a s}$ as a function of the source-detector distance $L$,
where $\sin^{2}2\theta_{ a s}\equiv \mathcal{C}\sin^{2}(2\theta_{e s})+(1-\mathcal{C})\sin^{2}(2\theta_{\mu s})$.
The initial neutrino beam flux at $L=0$ consists
of electron and muon neutrinos. Here we assume the mean energy
of neutrino beam $\bar E_\nu\sim\, 1\,{\rm GeV}$.
The dashed line represents the neutrino oscillations for the case that the neutrino beam has a very sharp energy distribution around $\bar E_\nu$.
The solid line represents the neutrino oscillations for the case that the neutrino beam has a Gaussian energy distribution with the mean energy
$\bar E_\nu$ and spreading width $\sigma=0.1\bar E_\nu$. }\label{oscf}
\end{figure}

Based on the oscillation from active neutrinos to sterile neutrinos
(\ref{flux3}) in the short baseline experiments, we estimate the rate of
generating circular polarization of laser photons at the source-detector
distance $L$
 \begin{equation}\label{rate}
    R_{_V}(L)\approx \frac{1}{k} \sigma_{\rm laser}\, f_{\rm pulse} \,  \, \tau_{\rm pulse}\Delta V (L),
\end{equation}
where $\tau_{\rm pulse}$ is the time duration of a laser pulse, the effective area of photon-neutrino interaction is represented by the laser-beam size $\sigma_{\rm laser}$ being smaller than the
neutrino-beam size $\Delta d$, and the laser repetition rate $f_{\rm pulse}$ is the number of laser pulses per second. To have more efficiency, we assume that laser and neutrino beams are synchronized and the $f_{\rm pulse}$ is equal to the repetition rate of neutrino beam $f_{\rm bunch}$, which is the number of neutrino bunches per second.
Using Eqs. (\ref{flux3}-\ref{rate}), the rate of generating circular polarization of laser photons is given as a function of the source-detector distance $L$
\begin{eqnarray}
R_{_V}(L) &=&
 R_{_V}(0)\left[1-\big(\mathcal{C}\sin^{2}(2\theta_{e s})+(1-\mathcal{C})\sin^{2}(2\theta_{\mu s})\big)\sin^{2}(\frac{\pi\,L}{L_{\rm osc}})\right],
  \label{circular2}
\end{eqnarray}
where
\begin{eqnarray}
L_{\rm osc} = \frac{4\pi \bar{E}_\nu}{\Delta m^2};\quad R_{_V}(0) =\frac{1}{k} \sigma_{\rm laser}\, f_{\rm pulse} \,  \, \tau_{\rm pulse}\Delta V (L=0). \label{osc}
\end{eqnarray}
If we consider the mean energy of neutrino beam $\sim \,1 \,{\rm GeV}$ and
$\Delta m^2\,\sim 1\,{\rm eV}^2$, the oscillation length $L_{\rm osc}\simeq 2.5 \,{\rm Km}$.
As shown in Fig.(\ref{oscf}), the oscillation has its minimum at
$L=nL_{\rm osc}/2$
and maximum  $L=(n+1)L_{\rm osc}/2$, $n=1,2,3,\cdot\cdot\cdot$. Measuring
the value $L=L_{\rm osc}/2$, one can determine the value of
$\Delta m^2$ the squared-mass
difference of active and sterile neutrinos, thus approximately obtain the sterile neutrino mass
$m_{\nu_s}$ for $m_{\nu_s}\gg m_{\nu_i}$.

As discussed in Refs.~\cite{xue1,roh1}, with a neutrino beam $\bar F_\nu\sim 10^{4}\,{\rm GeV}\,{\rm cm}^{-2}\,{\rm sec}^{-1}$ (see for example \cite{T2K})
and a linearly polarized laser beam of energy $k\sim $eV and power $\bar{P}_{\rm laser}\simeq 10$MW, the rate of generating circularly polarized photons $R_{_{\rm V}}(L=0) \sim 1/{\rm sec}$ ($\sim 9\times10^4/{\rm day}$).
This value $R_V(L)$ depends on the compositions (${\mathcal C}$) of the initial neutrino beam.
For a pure muon (electron) neutrino beam at $L=0$,
we have $R_{_V}(L)/R_{_V}(0)-1 < 0.05\, (0.13)$ (see Eq.~(\ref{data})). For
the case of mixing muon-electron neutrino beam
$R_{_V}(L)/R_{_V}(0)-1 <  0.13\, \mathcal{C}  + 0.13\, (1-\mathcal{C})<0.13 $.
If we use an electron neutrino beam with the mean energy
$\sim {\rm GeV}$ ($\Delta m^2\sim 1 {\rm eV}^2$), the rate of generating
circularly polarized photons decreases about $13\%$ at $L=L_{\rm osc}/2 $,
i.e., $R_{_V}(L_{\rm osc}/2) \sim 0.87/{\rm sec}$,
in comparison with $R_{_{\rm V}}(L=0) \sim 1/{\rm sec}$ at $L=0$. This
variation should be detectable so as to determine
$\Delta m^2$, $\sin^{2}(2\theta_{e s})$ and $\sin^{2}(2\theta_{\mu s})$ by
Eq.~(\ref{circular2}). Considering the damping of neutrino oscillations
with the distance $L$ (see Fig.~\ref{oscf}), one should appropriately
measure the locations of first three minimal (or maximal)
values of $R_V(L)$ at $L=L_{\rm osc}/2, 3\,L_{\rm osc}/2$ and
$5\,L_{\rm osc}/2$ so that $\Delta m^2$, $\sin^{2}(2\theta_{e s})$ and $\sin^{2}(2\theta_{\mu s})$ can be determined.

\section{Conclusion and remarks}
In this Letter, we give a brief discussion on the possible relation of
the anomalies observed in short baseline experiments to the oscillation
between active and sterile neutrinos and other alternative scenarios which explains short baseline neutrino anomaly. In order to measure constrains of the oscillation
amplitudes and squared mass difference in short baseline experiments and alternative scenarios,
we propose the interacting of laser beams with the neutrino beam and
measuring the generated circular polarization of laser photons.
This bases on the result \cite{xue1}
that the circular polarization of laser photons is generated by the collision
of laser and neutrino beams. This phenomenon was also considered \cite{roh1}
for obtaining the power spectrum ${\mathcal C}_l^V$ of the circular
polarization of CMB photons. As discussed in Ref.~\cite{xue1}, the rate of
generating the circular polarization of laser photons should be large enough for experimental measurements. While all active flavor neutrinos have approximately the same amplitude, number of events for different baseline are the same in there is not any active-sterile mixing. Thus with this method can test active-sterile neutrino mixing versus all other alternative without consideration of sterile neutrino such as quantum decoherence and CPT or Lorentz symmetry violation. Moreover if there is any active-sterile neutrino mixing it can be constrained via this experiment. In conclusion, this proposal should add
a valuable information for understanding the physics of sterile neutrinos and their oscillations with active neutrinos versus othe alternatives methods.


\end{document}